\documentclass[preprint,12pt]{elsarticle}

\usepackage{graphicx}
\usepackage{booktabs}
\usepackage{amsmath}
\usepackage{multirow}
\usepackage{hyperref}
\usepackage{float}

\journal{Journal of Biomedical Informatics}

\begin{document}
\begin{frontmatter}

\title{Grounded Multimodal Retrieval-Augmented Drafting of Radiology Impressions Using Case-Based Similarity Search}

\author{Himadri Sekhar Samanta}
\address{Independent AI Researcher, Austin, Texas, USA}

\begin{abstract}
Automated radiology report generation has gained increasing attention with the rise of deep learning and large language models. However, fully generative approaches often suffer from hallucinations and lack clinical grounding, limiting their reliability in real-world workflows. In this study, we propose a multimodal retrieval-augmented generation (RAG) system for grounded drafting of chest radiograph impressions. The system combines contrastive image-text embeddings, case-based similarity retrieval, and citation-constrained draft generation to ensure factual alignment with historical radiology reports.

A curated subset of the MIMIC-CXR dataset was used to construct a multimodal retrieval database. Image embeddings were generated using CLIP encoders, while textual embeddings were derived from structured impression sections. A fusion similarity framework was implemented using FAISS indexing for scalable nearest-neighbor retrieval. Retrieved cases were used to construct grounded prompts for draft impression generation, with safety mechanisms enforcing citation coverage and confidence-based refusal.

Experimental results demonstrate that multimodal fusion significantly improves retrieval performance compared to image-only retrieval, achieving Recall@5 above 0.95 on clinically relevant findings. The grounded drafting pipeline produces interpretable outputs with explicit citation traceability, enabling improved trustworthiness compared to conventional generative approaches. This work highlights the potential of retrieval-augmented multimodal systems for reliable clinical decision support and radiology workflow augmentation.
\end{abstract}

\begin{keyword}
Radiology report generation \sep multimodal retrieval \sep retrieval-augmented generation \sep chest X-ray \sep medical imaging \sep explainable AI
\end{keyword}

\end{frontmatter}


\section{Introduction}
Artificial intelligence has shown promise in assisting radiologists with image interpretation and report drafting. Deep learning has substantially improved performance in medical image analysis tasks \cite{litjens2017survey,miotto2018deep}, including thoracic imaging, where chest radiographs are among the most widely performed diagnostic studies \cite{rajpurkar2017chexnet,irvin2019chexpert}. More recently, large language models and multimodal systems have enabled automated clinical text generation, but they remain vulnerable to hallucinations, unsupported claims, and poor calibration in safety-critical settings \cite{openai2023gpt4,kelly2019key}. These limitations are especially problematic in radiology, where factual consistency and traceability are essential.

Retrieval-augmented generation (RAG) offers an alternative by constraining generation to retrieved evidence \cite{lewis2020rag,gao2024ragsurvey}. Instead of producing free-form text solely from model priors, RAG systems retrieve semantically similar historical examples and condition generation on verifiable evidence. Such case-based reasoning aligns closely with radiology practice, where prior studies and similar historical cases frequently inform interpretation. In radiology, both the chest X-ray image and the report impression are informative: the image captures anatomical and pathological appearance, while the report impression encodes concise clinical interpretation.

In this work, we present a grounded multimodal retrieval-augmented system for chest radiograph impression drafting. The proposed framework combines image and report embeddings, performs late fusion for similarity retrieval, and indexes the resulting vectors with FAISS for efficient nearest-neighbor search \cite{johnson2021faiss}. Retrieved cases are then used to generate grounded draft impressions with explicit case citations. To improve safety, the system applies a confidence threshold and refuses generation when similarity is insufficient. We additionally deploy the pipeline as a FastAPI service and package it for reproducible Docker-based inference.

The main contributions of this work are as follows:
\begin{itemize}
    \item We construct a clean multimodal chest X-ray dataset subset from MIMIC-CXR with aligned image paths, study identifiers, and extracted impression text suitable for retrieval and drafting experiments \cite{johnson2019mimiccxr}.
    \item We demonstrate that multimodal fusion of image and text embeddings markedly improves retrieval performance over image-only retrieval.
    \item We introduce a grounded drafting mechanism with explicit case citations and confidence-based refusal to reduce unsupported output generation.
    \item We provide an end-to-end deployable implementation, including a REST API and containerized execution environment, illustrating practical translation beyond notebook experimentation.
\end{itemize}

\section{Related Work}
Automated radiology report generation has been studied using encoder--decoder models, transformer-based image captioning systems, and multimodal vision-language architectures. While these approaches have shown promising fluency, they often struggle with factual consistency and hallucination. Chest X-ray modeling has particularly benefited from large public datasets and weakly supervised labeling pipelines, such as CheXpert \cite{irvin2019chexpert}. Contrastive language-image pretraining further demonstrated that aligned multimodal embeddings can support strong cross-modal representations for retrieval and transfer learning \cite{radford2021clip}. Medical-domain adaptations, including BioViL and XrayBERT, highlight the growing role of clinically aligned vision-language pretraining \cite{delbrouck2022biovil,boecking2022xraybert}.

In parallel, retrieval-based and case-based reasoning methods have been explored in radiology because they naturally align with clinical practice and can improve interpretability. Efficient similarity search frameworks such as FAISS support practical deployment of large embedding databases \cite{johnson2021faiss}. More recently, RAG pipelines have been proposed as a way to improve factuality and evidence grounding in language generation \cite{lewis2020rag,gao2024ragsurvey,tang2023medrag,chen2023clinicalrag}. However, relatively few systems combine radiology images and report text within a unified retrieval architecture that also includes citation verification and confidence-based refusal. Our work addresses this gap by integrating multimodal retrieval, grounded drafting, safety gating, and deployable serving in a single workflow.

Explainability and deployment realism are also central concerns in clinical AI \cite{holzinger2017ai,esteva2019guide,topol2019high}. Prior work has emphasized that successful healthcare AI must not only achieve strong offline metrics but also provide interpretable outputs, calibrated uncertainty, and operational pathways to translation \cite{kelly2019key,zhou2023foundation}. Our design choices are motivated by these principles.

\section{Dataset}
We used a curated subset of the MIMIC-CXR dataset and the MIMIC-CXR-JPG release \cite{johnson2019mimiccxr}. The development process intentionally avoided downloading the full multi-terabyte dataset. Instead, lightweight metadata tables and the compressed report archive were first downloaded from PhysioNet under credentialed access. A reproducible subset of 2,000 studies was sampled using metadata mappings between subject identifiers, study identifiers, and image identifiers. After expansion to DICOM-level rows, the initial subset contained 3,301 entries.

To construct a modeling-ready image dataset, the study manifest was then joined to MIMIC-CXR-JPG metadata in order to retain only studies with available JPG images. This yielded 3,149 rows with corresponding image files. The \texttt{IMPRESSION} section was extracted from each report using rule-based section parsing, as the impression provides the most concise clinically relevant summary for retrieval and draft generation. Rows without a valid impression section were removed, resulting in a final clean multimodal dataset of 2,696 image--impression pairs. Table~\ref{tab:dataset_summary} summarizes the dataset construction process.

\begin{table}[ht]
\centering
\caption{Dataset construction summary.}
\label{tab:dataset_summary}
\begin{tabular}{lr}
\toprule
Stage & Count \\
\midrule
Sampled studies & 2000 \\
DICOM rows in initial manifest & 3301 \\
Rows with available JPG images & 3149 \\
Rows with valid impression text & 2696 \\
Final clean multimodal dataset & 2696 \\
Positive-label evaluation subset & 839 \\
\bottomrule
\end{tabular}
\end{table}

\section{Methods}

\subsection{Image representation learning}
Each chest X-ray image was encoded using the CLIP ViT-B/32 image encoder \cite{radford2021clip}. Images were loaded as RGB inputs, preprocessed using the CLIP processor, and converted into 512-dimensional embeddings. All image embeddings were L2-normalized to facilitate cosine similarity retrieval through inner-product search.

\subsection{Text representation learning}
For each study, the extracted report impression was encoded using the CLIP text encoder. Texts were tokenized with padding and truncation, embedded into the same 512-dimensional space, and L2-normalized. Because both encoders are aligned through contrastive pretraining, the image and text vectors are directly comparable.

\subsection{Multimodal fusion}
Late fusion was used to combine image and text embeddings:
\[
\mathbf{e}_{fusion} = \alpha \mathbf{e}_{image} + (1-\alpha)\mathbf{e}_{text},
\]
where $\alpha \in [0,1]$ controls the contribution of the visual branch. The fused vectors were normalized after combination. We evaluated multiple values of $\alpha$ and selected the setting with the best retrieval performance on the development subset. The best-performing setting was $\alpha = 0.5$. Figure~\ref{fig:alphaablation} summarizes the fusion-weight ablation.

\subsection{Retrieval system}
All normalized fused embeddings were indexed using FAISS with an inner-product similarity index \cite{johnson2021faiss}. Because the vectors were normalized, inner product is equivalent to cosine similarity. At inference time, the system computes an embedding for the query image and retrieves the top-$K$ most similar cases. For deployment, we used an image-only query against the image-side index to avoid a modality mismatch between query-time inputs and indexed vectors, while multimodal fusion was retained for offline evaluation and ablation.

\subsection{Grounded draft generation}
For each query, the top-$K$ retrieved cases were used as supporting evidence. Retrieved report impressions were converted into short evidence snippets, and a grounded draft impression was produced either by a lightweight language model or by a deterministic evidence-based summarizer. The deterministic fallback ensures that the draft remains constrained to retrieved evidence and can be used when the language model output is malformed or insufficiently grounded.

\subsection{Citation verification}
Each draft output was required to reference retrieved evidence using explicit case identifiers (e.g., [Case 1], [Case 2]). Citation coverage was computed as the fraction of expected case markers present in the final draft. Missing citations were recorded for error analysis. This provides a simple but effective mechanism to track evidence attribution.

\subsection{Confidence-based refusal}
The similarity score of the top retrieved case was used as a confidence signal. When the top-1 similarity score fell below a predefined threshold, the system refused to generate a report draft and instead returned a structured refusal response. This mechanism prevents low-confidence or out-of-distribution inputs from producing unsupported clinical text.

\subsection{Deployment}
The final system was deployed as a FastAPI-based REST service with two primary endpoints: \texttt{/health} for health checks and \texttt{/predict} for inference. The service returns prediction status, confidence score, latency, generated draft, and retrieved case identifiers. For reproducibility, the system was packaged in Docker with a version-controlled dependency file.

\section{Experiments}

\subsection{Retrieval evaluation}
We evaluated retrieval quality using Recall@$K$ with $K \in \{1,5,10\}$. Relevance was defined using CheXpert-derived pathology labels \cite{irvin2019chexpert}. A query was considered successful if at least one retrieved case among the top-$K$ shared the same positive clinical label as the query. We additionally compared image-only retrieval against multimodal fusion retrieval.

\subsection{Fusion weight study}
To understand the contribution of image and text signals, we performed an alpha sweep over fusion weights and measured Recall@5. The best setting was retained for subsequent experiments.

\subsection{Safety evaluation}
We evaluated the deployed RAG layer using refusal rate, average top-1 retrieval similarity, and average citation coverage. Refusal rate quantifies how often the safety policy suppresses generation; citation coverage measures the extent to which the final report explicitly references retrieved evidence.

\section{Results}

\subsection{Retrieval performance}
\begin{figure}[H]
\centering
\includegraphics[width=0.72\textwidth]{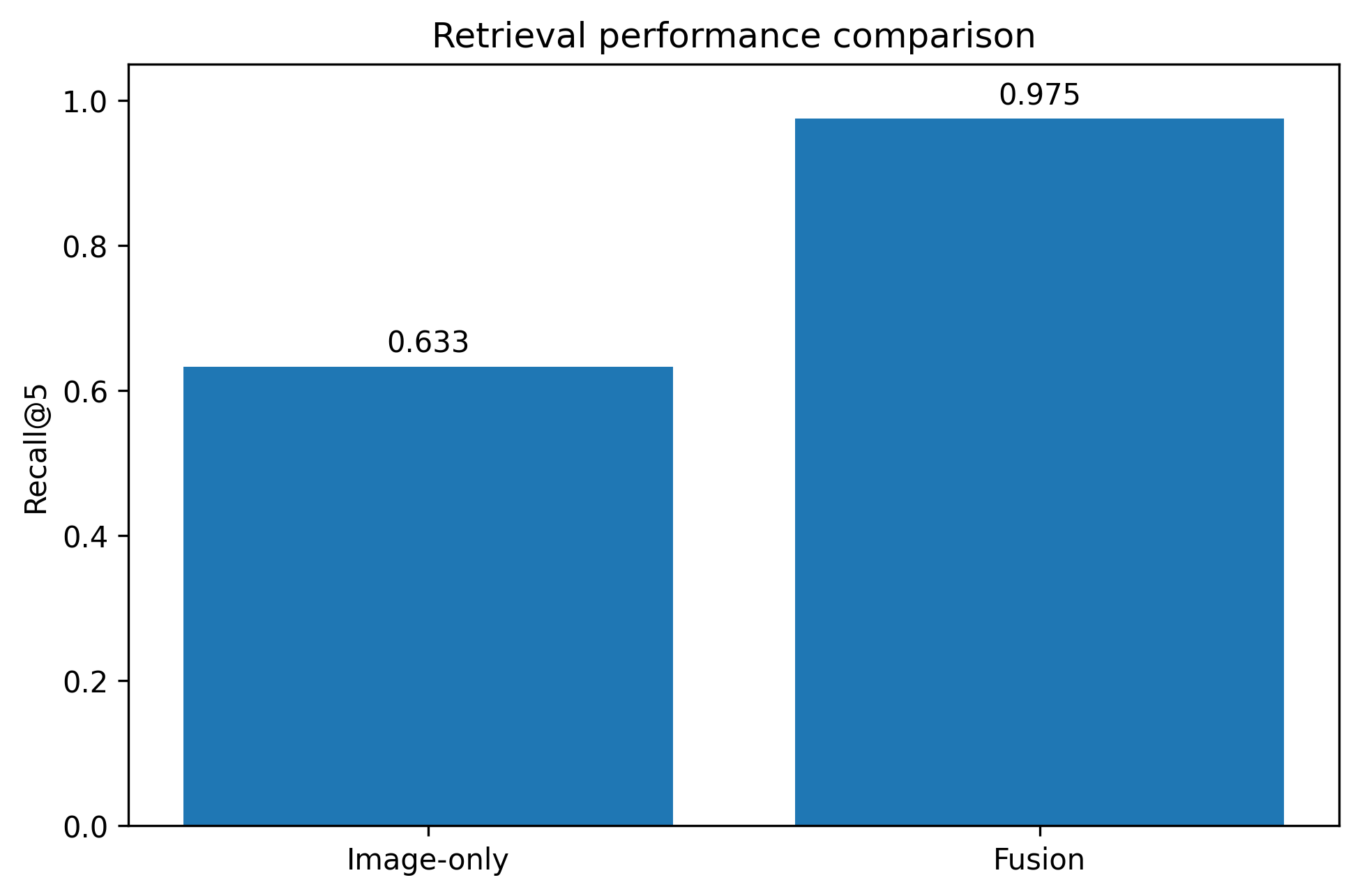}
\caption{Recall@5 comparison between image-only retrieval and multimodal fusion.}
\label{fig:recall5compare}
\end{figure}

\begin{figure}[H]
\centering
\includegraphics[width=0.72\textwidth]{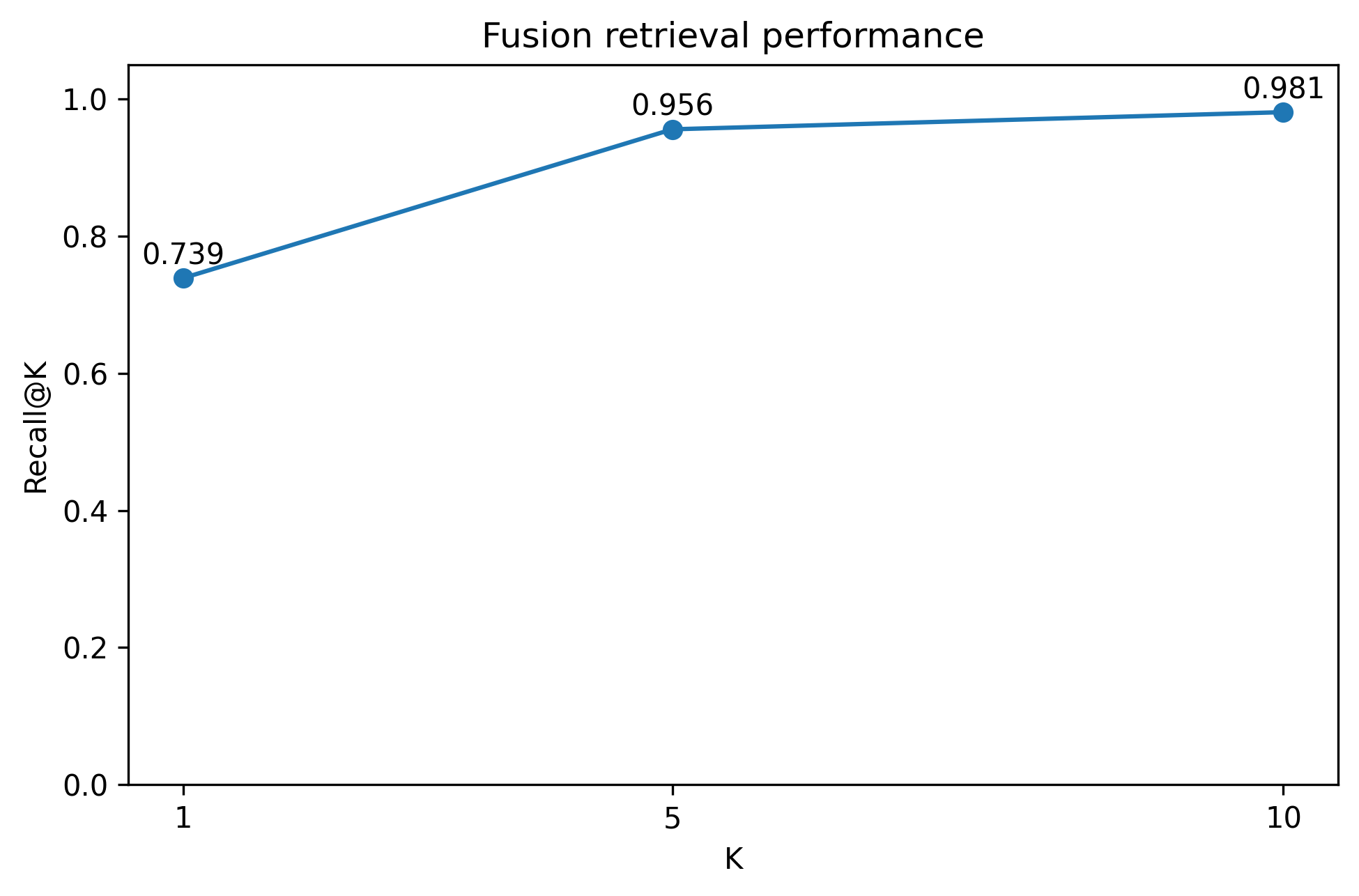}
\caption{Fusion retrieval performance across different values of $K$.}
\label{fig:recallatk}
\end{figure}

\begin{figure}[H]
\centering
\includegraphics[width=0.72\textwidth]{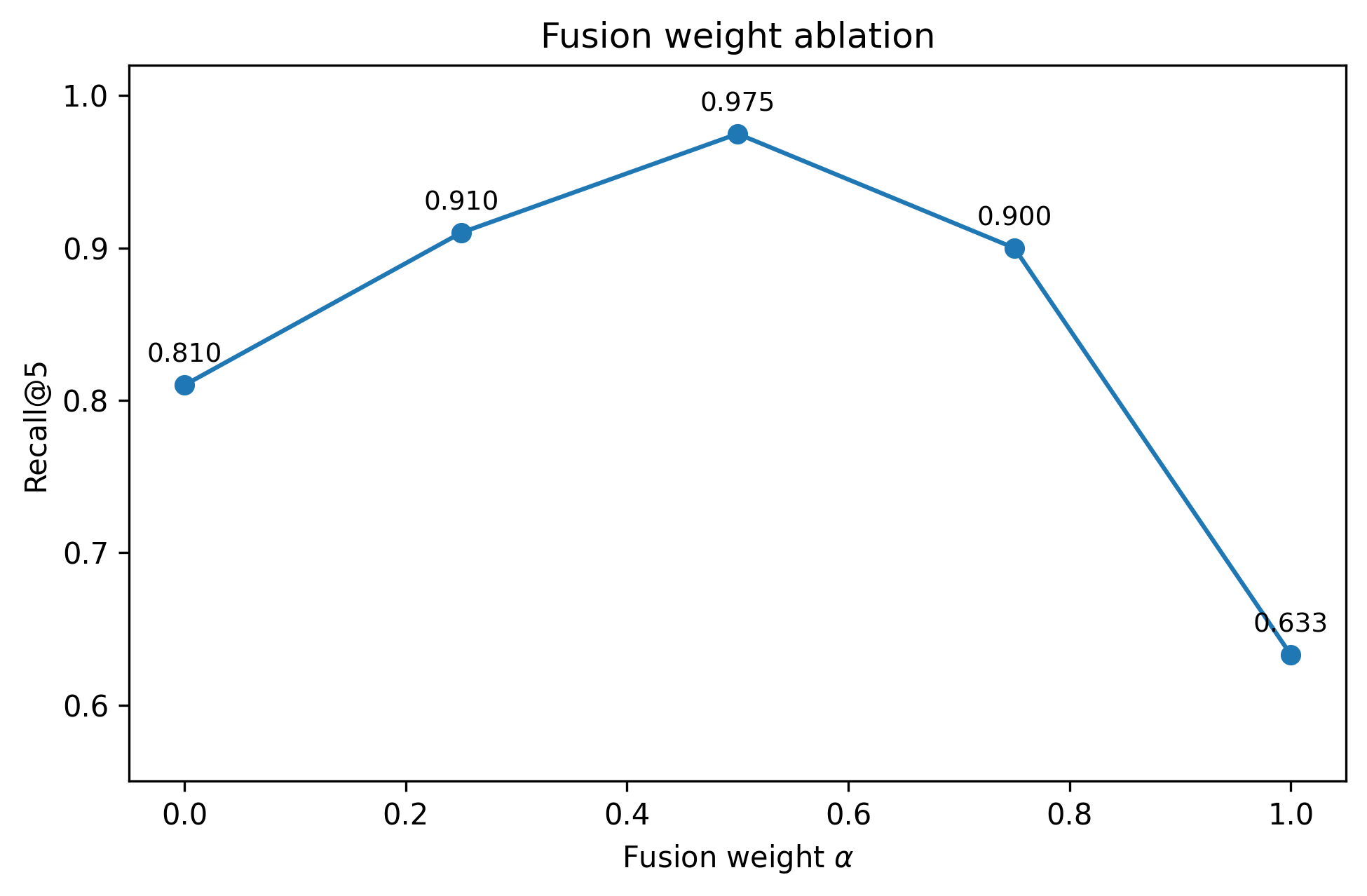}
\caption{Fusion-weight ablation showing peak retrieval performance near $\alpha=0.5$.}
\label{fig:alphaablation}
\end{figure}

Multimodal fusion substantially improved retrieval performance over image-only retrieval. Table~\ref{tab:retrieval_results} summarizes the main retrieval metrics, while Figures~\ref{fig:recall5compare}, \ref{fig:recallatk}, and \ref{fig:alphaablation} provide visual summaries.

\begin{table}[ht]
\centering
\caption{Retrieval performance comparison.}
\label{tab:retrieval_results}
\begin{tabular}{lccc}
\toprule
Method & Recall@1 & Recall@5 & Recall@10 \\
\midrule
Image-only & --- & 0.633 & --- \\
Fusion ($\alpha=0.5$) & 0.739 & 0.956 & 0.981 \\
Best fusion setting & --- & 0.975 & --- \\
\bottomrule
\end{tabular}
\end{table}

Image-only retrieval achieved Recall@5 of 0.633, indicating that visual similarity alone captures some clinically useful signal but misses a substantial fraction of relevant cases. In contrast, multimodal fusion achieved Recall@5 of 0.975 at the best fusion setting and Recall@5 of 0.956 on the positive-case evaluation subset. These results indicate that report semantics provide complementary information beyond image appearance alone.

\subsection{Safety and reliability}
\begin{figure}[H]
\centering
\includegraphics[width=0.75\textwidth]{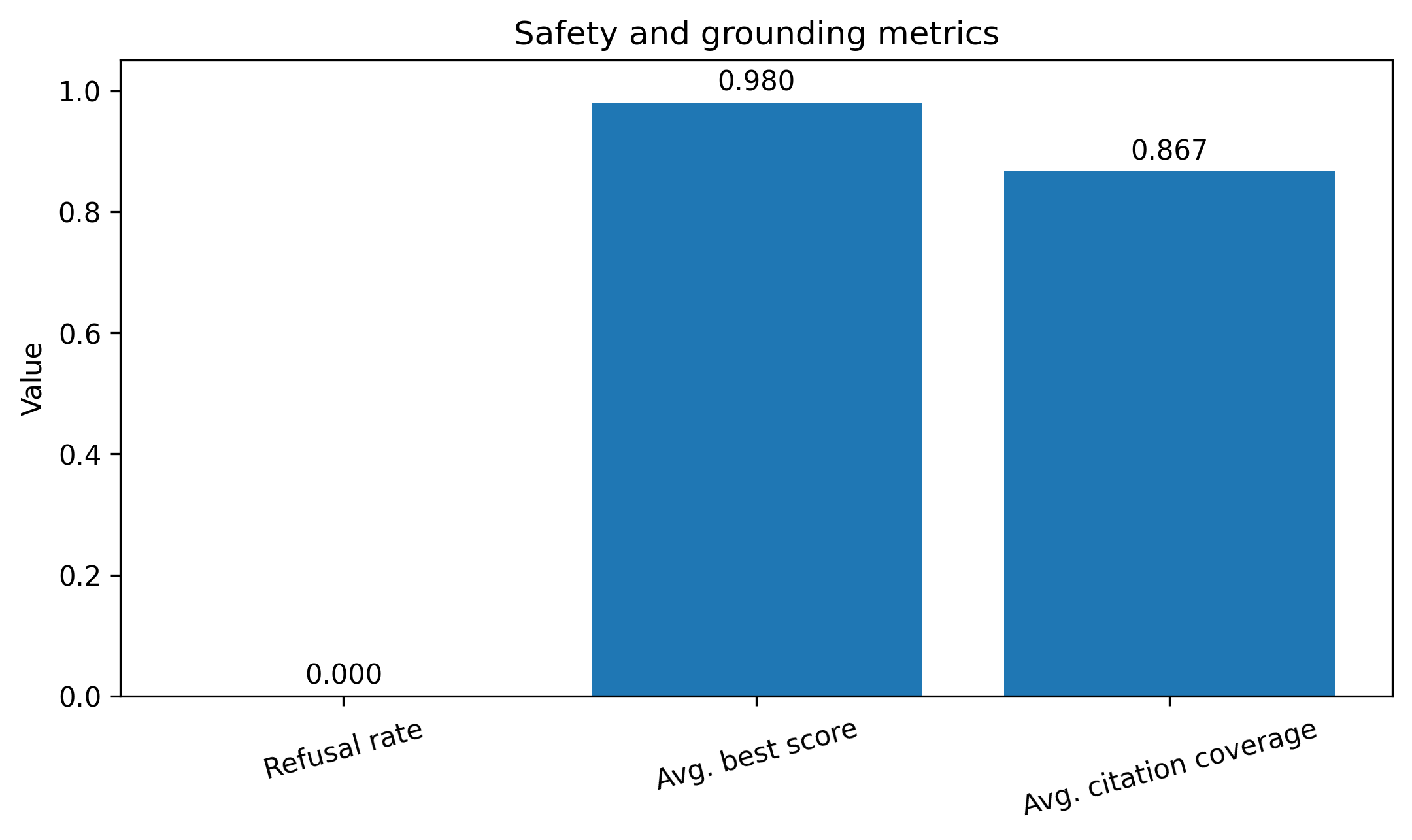}
\caption{Safety and grounding metrics: refusal rate, average best retrieval score, and average citation coverage.}
\label{fig:safety}
\end{figure}

The deployed RAG system demonstrated strong safety metrics. As shown in Table~\ref{tab:safety} and Figure~\ref{fig:safety}, the refusal rate on the internal evaluation set was 0.000, average top-1 retrieval similarity was 0.980, and average citation coverage was 0.867. These results indicate that most in-distribution chest X-ray inputs retrieve strong matches and that the majority of generated statements remain explicitly grounded in cited evidence.

\begin{table}[ht]
\centering
\caption{Safety and grounding metrics.}
\label{tab:safety}
\begin{tabular}{lc}
\toprule
Metric & Value \\
\midrule
Refusal rate & 0.000 \\
Average top-1 retrieval score & 0.980 \\
Average citation coverage & 0.867 \\
Best fusion weight $\alpha$ & 0.5 \\
\bottomrule
\end{tabular}
\end{table}

\subsection{Qualitative example}
Table~\ref{tab:qualitative_example} presents a representative case illustrating the proposed grounded drafting approach.

\begin{table}[ht]
\centering
\caption{Representative qualitative example of grounded drafting.}
\label{tab:qualitative_example}
\begin{tabular}{p{0.21\linewidth} p{0.69\linewidth}}
\toprule
Component & Example \\
\midrule
Query finding & Mild bibasilar atelectatic change on frontal chest radiograph \\
Retrieved Case 1 & Bibasilar atelectasis. Otherwise, no acute cardiopulmonary abnormality. \\
Retrieved Case 2 & Mild bibasilar atelectasis. No other acute findings. \\
Retrieved Case 3 & Mild bibasilar atelectasis. Otherwise, no acute cardiopulmonary process. \\
Generated draft & Mild bibasilar atelectasis. [Case 1][Case 2] No acute cardiopulmonary abnormality. [Case 1][Case 3] \\
\bottomrule
\end{tabular}
\end{table}

\subsection{Architecture overview}
\begin{figure}[H]
\centering
\includegraphics[width=\textwidth]{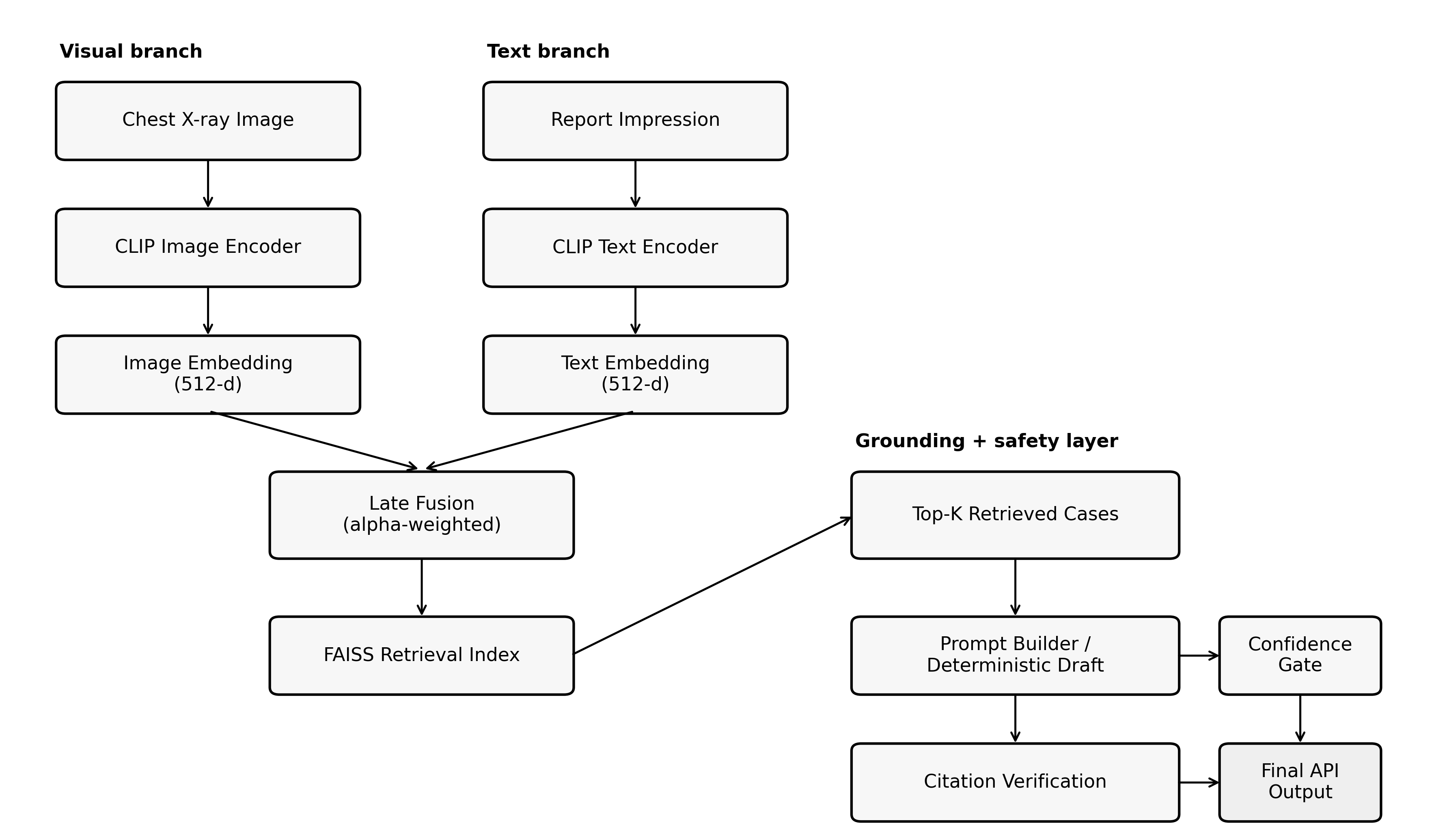}
\caption{System architecture for grounded multimodal retrieval-augmented radiology drafting.}
\label{fig:architecture}
\end{figure}

Figure~\ref{fig:architecture} illustrates the complete multimodal RAG pipeline used in this work, from image and text encoding to retrieval, grounded drafting, citation verification, confidence gating, and API output.

\section{Error Analysis}
The main error patterns observed in development fell into three categories. First, image-only retrieval sometimes returned anatomically similar but clinically different cases, especially when subtle findings required report semantics for disambiguation. Second, the lightweight language model occasionally copied evidence markers or repeated retrieved content verbatim rather than producing a concise synthesized impression. Third, some drafts omitted one or more expected citations, reducing citation coverage. These issues motivated the use of a deterministic evidence-based fallback and explicit citation verification.

We also verified the behavior of the confidence-based refusal mechanism on out-of-domain images. When a non-chest photograph was submitted, the system returned a low similarity score and correctly refused to generate a draft under the default clinical threshold. Lowering the threshold enabled generation in demo mode, but this setting is not recommended for clinical use.

\section{Discussion}
This study demonstrates that multimodal retrieval is a strong foundation for safer radiology draft generation. The retrieval results suggest that the textual impression signal contributes critical clinical information that image-only representations do not fully capture. The strong Recall@5 and Recall@10 results indicate that the fused embedding space successfully organizes studies according to clinically meaningful similarity.

The RAG layer provides an important practical advantage: rather than asking a model to generate a report from scratch, the system constrains drafting to retrieved, previously observed evidence. This substantially improves interpretability. In addition, confidence gating provides a pragmatic safety control by allowing the system to abstain when retrieval evidence is weak. Together, these components align with real clinical requirements for traceability and conservative behavior \cite{holzinger2017ai,kelly2019key}.

The deployed API and Docker packaging are also significant from an applied perspective. Many academic systems stop at offline metrics, whereas our implementation demonstrates a path toward reproducible and portable deployment. This increases the work's relevance to both academic and industry audiences \cite{esteva2019guide,topol2019high,zhou2023foundation}.

\section{Limitations}
This work has several limitations. First, experiments were conducted on a curated subset rather than the full MIMIC-CXR dataset, which may limit generalization. Second, the report drafting component relied on simple prompt-based or deterministic summarization rather than a radiology-specialized large language model. Third, evaluation emphasized retrieval and evidence-grounding metrics, but did not yet include formal radiologist review of generated outputs. Finally, citation coverage is a useful but imperfect proxy for factual correctness; future work should incorporate stronger faithfulness evaluation, larger-scale validation, and external benchmarking \cite{wang2023multimodal,karargyris2021radimagenet}.

\section{Conclusion}
We presented a grounded multimodal retrieval-augmented radiology co-pilot for chest X-ray impression drafting. The proposed framework combines image and report embeddings, uses FAISS for efficient case retrieval, and generates citation-grounded drafts with a confidence-based refusal policy. Experiments on a curated MIMIC-CXR subset show that multimodal fusion greatly improves retrieval performance over image-only baselines and that the deployed system achieves strong retrieval confidence and citation coverage. These results support the use of multimodal retrieval and explicit evidence grounding as promising directions for safer and more interpretable AI assistance in radiology workflows.

\bibliographystyle{elsarticle-num}
\bibliography{references}

\end{document}